# Room-Temperature High-Purity Single Photon Emission from Carbon-Doped Boron Nitride Thin Films

Arka Chatterjee[1], Abhijit Biswas[2], Addis S. Fuhr[3], Tanguy Terlier[4], Bobby G. Sumpter[3], Pulickel M. Ajayan[2], Igor Aharonovich[5,6], and Shengxi Huang[1,7*]

[1] *Department of Electrical and Computer Engineering, Rice University, Houston, TX, 77005, USA*

[2] *Department of Materials Science and Nanoengineering, Rice University, Houston, TX, 77005, USA*

[3] *Center for Nanophase Materials Sciences, Oak Ridge National Laboratory, Oak Ridge, TN, 37831, USA*

[4] *SIMS laboratory, Shared Equipment Authority, Rice University, Houston, TX, 77005, USA*

[5] *School of Mathematical and Physical Sciences, University of Technology Sydney, Ultimo, New South Wales 2007, Australia*

[6] *ARC Centre of Excellence for Transformative Meta-Optical Systems, University of Technology Sydney, Ultimo, New South Wales 2007, Australia*

[7] *Rice Advanced Materials Institute, Rice University, Houston, TX, 77005, USA*

*Corresponding Author. E-mail: shengxi.huang@rice.edu*

## Abstract

Hexagonal boron nitride (h-BN) has emerged as an excellent host material for generating room temperature single photons exhibiting high brightness and spin-photon entanglement. However, challenges in improving purity, stability, and scalability limit its use in quantum technologies. Here, we demonstrate highly pure and stable single photon emitters (SPEs) in h-BN by directly growing carbon-doped, centimeter-scale h-BN thin films using the pulsed laser deposition (PLD) method. These SPEs exhibit room temperature operation with polarized emission, achieving a $g^{(2)}(0)$ value of 0.015, which is among the lowest reported for room temperature SPEs and the lowest achieved for h-BN SPEs. It also exhibits high brightness (~0.5 million counts/second), remarkable stability during continuous operation (>15 minutes), and a Debye-Waller factor of 45%. First-principles calculations reveal unique carbon defects responsible for these properties, enabled by PLD's low-temperature synthesis and in-situ doping. Our results demonstrate an effective method for large-scale production of high-purity, stable SPEs in h-BN, enabling robust quantum optical sources for various quantum applications.

# Teaser

Carbon-doped h-BN thin film shows single photon emission with record high purity at room temperature.

# INTRODUCTION

Solid-state single-photon emitters (SPEs) are indispensable components for advancing quantum technologies, including secure communication, computing with quantum advantage, and precise metrology (*1–3*). Hexagonal boron nitride (h-BN) has recently emerged as a promising candidate to host SPEs, demonstrating room temperature, bright, and polarized single photon emission (*4–7*). Unlike narrow-band gap semiconductors, where SPEs are typically attributed to shallow defects and strain, emitters in h-BN are mainly associated with crystallographic defects such as antisites, substitutions, vacancies, or a combination of them with their electronic levels deep within its large band gap (~ 6 eV) (*8, 9*). Additionally, the excellent chemical and thermal robustness of h-BN ensures long-term stable operation of these SPEs. Several strategies have been adopted to create atomic defects in h-BN, which include ion irradiation (*10*), focused ion beam (FIB) treatment (*11*), plasma treatment (*8*), and femtosecond laser ablation (*12*), etc. These techniques have proved effective in creating SPEs in h-BN. On the other hand, challenges exist, such as a broad spectral distribution (1.6–2.2 eV), low emitter density, and poor single photon purity, stability, and brightness (*8, 13, 14*).

So far, most studies have focused on h-BN prepared either through exfoliation or chemical vapor deposition (CVD). Exfoliation is a simple and efficient method for material preparation (*5, 8, 15*). While it offers superior crystal quality, large-scale exfoliation remains a major challenge, hindering the development of large-scale quantum photonic architectures. Recently, CVD has emerged as a popular choice for creating SPEs on a large scale (*4, 16, 17*). However, CVD often involves harsh growth conditions, such as high reaction temperatures, long growth times, and the use of toxic elements such as borazine and its by-products, making the process experimentally challenging and raising safety concerns (*18, 19*). Moreover, the CVD-grown h-BN films often require post-treatment processes such as high-temperature annealing in air and ultraviolet (UV) ozone processing to improve the purity of SPEs (*20, 21*). Alongside CVD, physical vapor deposition (PVD) offers an alternative approach for preparing large-scale thin films using pure physical methods, typically under high vacuum and lower growth temperatures. Among PVD techniques, pulsed laser deposition (PLD) has been successfully employed for the synthesis of large-scale h-BN (*22, 23*). Compared to CVD, PLD offers several advantages, such as a high growth rate, precise control over thickness and morphology, and lower growth temperatures (*24*). Besides enabling large-scale growth of thin films, PLD could also facilitate in-situ doping. This capability would be highly advantageous for creating single and pure defects in the material, addressing challenges related to the brightness, stability, and purity of SPEs. However, PLD has not yet been utilized to realize SPEs in h-BN so far.

Here, we demonstrate the directly-grown large-area, high-quality, centimeter-scale h-BN thin films using PLD that hosts highly pure, bright, and photostable SPEs. We in-situ doped h-BN thin films with carbon (C) since C-related defects in h-BN have recently drawn attention to creating SPEs with low spectral distribution (*14*, *25–28*). However, further improvement in the purity, stability, and brightness of these emitters remains a key challenge. Using our unique PLD synthesis and in-situ doping, we achieved a single photon purity of 98.5%, the highest reported for room temperature h-BN SPEs produced by any methods. Furthermore, the emitters exhibit exceptional stability (no spectral wondering for continuous excitation of 15 min) and remarkable brightness (saturation emission rate of 0.466 million counts/second). These exceptional properties make our SPEs highly advantageous for quantum applications. We also performed first-principles calculations to identify the nature of the defects responsible for our experimental observation. Our findings indicate that the emitters likely originate from complex C substitutional defects, which exhibit optical properties surpassing those of many other defect types. This defect structure is favored at low synthesis temperatures, which is unique to our PLD method. Our study successfully produces exceptional SPEs in h-BN, achieving record-high purity, stability, brightness, and Debye-Waller (DW) factor. Furthermore, our scalable samples offer the potential for integration with large-area on-chip systems, representing a major advancement in quantum technologies. Our theoretical analyses and insights will serve as a foundation for the future production of high-quality quantum emitters and materials.

## RESULTS

### Synthesis and characterization of the thin film

The synthesis process of the C-doped h-BN thin film is schematically illustrated in **Figure 1a,** along with the real-time photo taken during the growth (**Figure 1b**). The picture of the film (**Figure 1b**) further highlights the scalability of this method, demonstrating the centimeter-size of the C-doped h-BN thin films (~ 0.6 cm × 0.6 cm). Atomic force microscopy (AFM) reveals that the as-grown sample has a smooth surface with a surface roughness of less than 1 nm. To investigate the growth-induced incorporation of C in h-BN, X-ray photoelectron spectroscopy (XPS) is employed. Before measurements, the samples were gently etched with $Ar^+$ ions to remove the adventitious surface C. As shown in **Figures 1c-d** and supporting **Figure S1**, distinctive B–N bonding peaks are observed in elemental B 1s (at ~190.8 eV) and N 1s (at ~398.3 eV) core level scans, accompanied by π-plasmon peaks (~9.0 eV apart from the primary B–N peak), which are characteristic of h-BN (*22*, *23*). Additionally, both spectra exhibit an additional peak at 188.0 eV (near B 1s) and 399.5 eV (near N 1s), corresponding to B-C and N-C bonds, respectively (*27*). The peaks associated with the C-B bond (~282.67 eV) and the C-N bond (~286.58 eV) are also evident in the C 1s spectra (**Figure 1e**). The presence of these bonds suggests that C atoms effectively substitute both B and N atoms during growth, aligning well with prior studies in C-doped h-BN (*29–31*). The atomic percentage (at%) of C was measured to be 1.05 ± 0.16%, which is in good agreement with the C at% in the target (1 at%). Further investigation using time-of-flight secondary-ion mass spectrometry (TOF-SIMS) (**Figure S2**) confirms a uniform distribution of B, N, and C elements throughout the film thickness of about 32 nm, with C-B and C-N bonds observed across the sample.

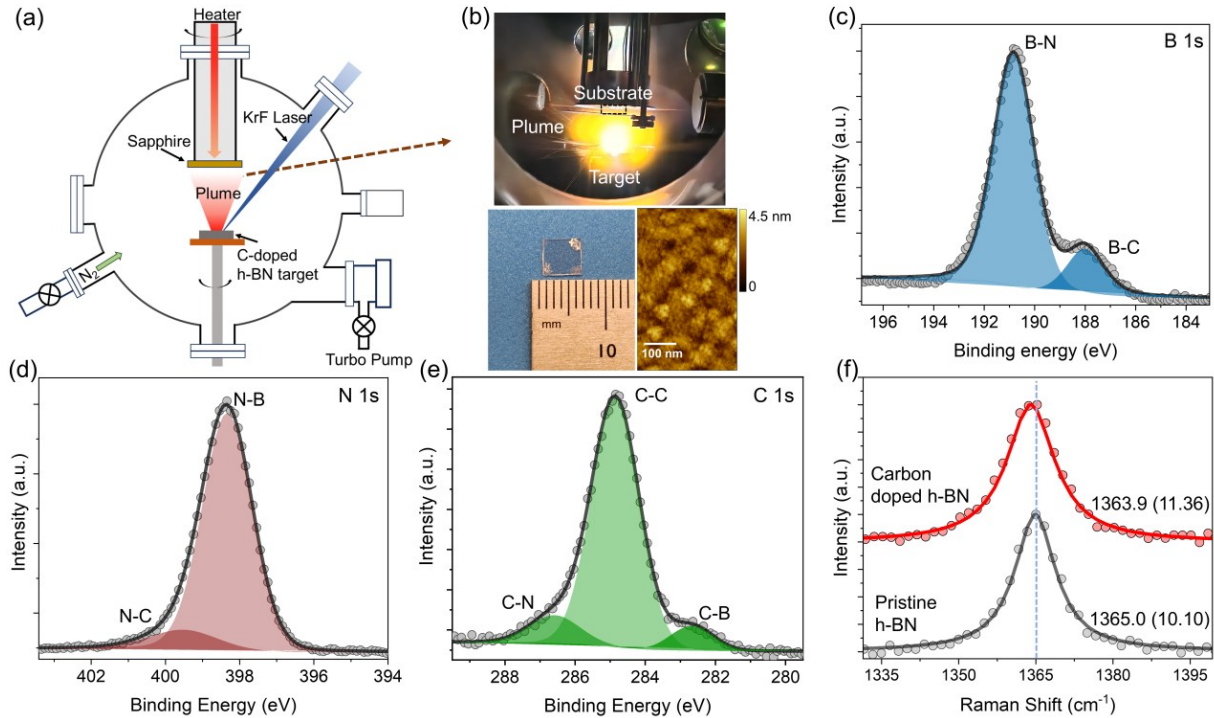

*Figure 1: Synthesis and characterization of the thin film. (a) Schematic of C-doped h-BN thin film growth using PLD. (b) Real-time plume generation during the laser ablation of the 1 at% C-*

doped h-BN target. Photo of the centimeter-scale thin film (left bottom panel) on the sapphire substrate and AFM image (right bottom panel) of the thin film. XPS spectra of (c) B 1s, (d) N 1s, and (e) C 1s core of C-doped h-BN thin film showing the B-N, B-C, and N-C peaks. (f) Raman spectra of pristine (black solid circles) and C-doped h-BN (red solid circles) thin films. The solid traces are the fit to the data.*

Next, we investigated and compared the crystalline quality of pristine and C-doped h-BN films using Raman spectroscopy. **Figure 1f** shows the Raman mode for pristine h-BN at 1365 cm$^{-1}$ corresponds to the $E_{2g}$ phonon. The full width at half maximum (FWHM) of this Raman mode is calculated to be 10.1 cm$^{-1}$, which is close to the value of single crystals (7-8 cm$^{-1}$) (*32*) and much lower than that of CVD-grown h-BN films, (17-40 cm$^{-1}$) (*16, 33*) demonstrating the high crystallinity of our material. After C-doping, a slight red shift of about 1.1 cm$^{-1}$ in the $E_{2g}$ phonon mode was observed. The FWHM is also found to increase to 11.4 cm$^{-1}$. These suggest the incorporation of C-related defects in the h-BN crystal, as observed in previous studies (*27*).

## Photophysical characterization

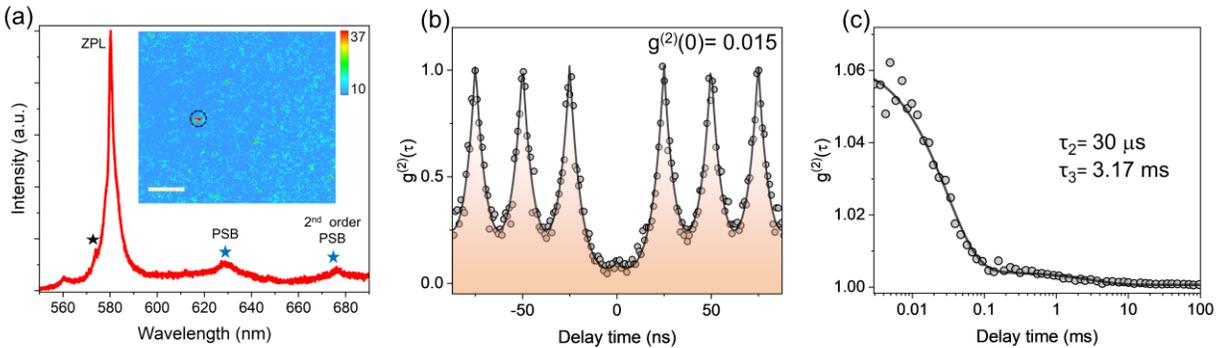

*Figure 2: Photophysical properties of the defects. (a) Room temperature PL spectrum of the emitter using 35 μW, CW 532 nm laser. Black asterisk around 573.8 nm denoting the Raman line. The PL mapping is shown in the inset, where a black circle marks the present emitter. The scale bar is 4 μm (b) Second-order autocorrelation functions, $g^{(2)}(\tau)$ for the emitter measured using 35 μW, 515 nm pulsed laser with a repetition rate of 40 MHz and a pulse width of 96 ps (without background correction). The solid black curve is fit to the data suggesting a $g^{(2)}(0)$ of 0.015. (c)*

*Second-order autocorrelation function, $g^{(2)}(\tau)$, measured at a longer time scale featuring bunching. The corresponding solid traces are fit to the experiment data.*

We then investigated the photophysical properties of our C-doped h-BN thin film. The photoluminescence (PL) spectrum of a bright and isolated emitter is shown in **Figure 2a.** The corresponding PL intensity map is also shown in the inset. The emitters are also found to be uniformly distributed in the sample (**Figure S3**). The spectrum features a narrow zero phonon line (ZPL) at 580.3 nm (2.136 eV) with FWHM of 3.0 nm (11 meV) along with first- and second-order phonon side band (PSB) at 629.2 (1.970 eV) and 677.0 nm (1.830 eV), respectively, originating from the interaction with longitudinal optical (LO) phonons (*34*). The energy detuning between the ZPL and PSB and second-order PSB are calculated to be 166 and 305 meV, respectively, which are found to be consistent with the previous C-doped h-BN reports (*25*). The sharp peak marked by an asterisk at ~ 573.8 nm in the PL spectrum originates from the Raman mode of C-doped h-BN. The PL spectrum is also fitted (**Figure S5a**) to determine the contribution of ZPL (45%), LA (30%), and LO (25%) phonons, which reveals lower phonon coupling to ZPL. Such narrow ZPL and lower phonon coupling are also less evident for h-BN quantum emitters around 580 nm. The DW factor (i.e., the fraction of light emitted into the ZPL) and the Huang-Rhys (S) factor (i.e., the average number of phonons emitted during an optical transition) are calculated to be 45% and 0.79, respectively. Higher DW and lower S factor signify that pure electronic transitions dominate, distinct from phonon-mediated transitions. The calculated DW factor of our emitter is found to be higher than the values of several popular SPEs, including G-centre in Si (~ 11% at 5 K) (*35*), SiC (~33% at 18 K) (*36*), and negatively charged nitrogen-vacancy (NV)⁻ center in diamond (~3% measured at room temperature) (*37*). As a control experiment, we also examined the pristine h-BN sample. However, no SPEs (**Figure S4**) were found, confirming that these SPEs originate from C rather than intrinsic defects in pristine h-BN.

Further, we investigated the single photon purity of the emitters using second-order photon correlation ($g^{(2)}(\tau)$) measurements on a Hanbury-Brown and Twiss (HBT) interferometer with pulsed laser excitations. **Figure 2b** shows the intensity autocorrelation measurements for the emitter. It is well known that the second-order correlation function $g^{(2)}(0)$ value below 0.5 at zero delays indicates single photon nature. Lower $g^{(2)}(0)$ suggests higher purity of SPEs. Our emitter shows a $g^{(2)}(0)$ of 0.015±0.002 without background correction, demonstrating superior single photon purity of 98.5±0.2% at room temperature. This $g^{(2)}(0)$ value is among the lowest reported for room temperature SPEs and the lowest achieved for h-BN SPEs produced by any method. Such a remarkably high purity is highly advantageous for various quantum technologies, such as quantum key distribution (*38*). Furthermore, the excited state lifetime of the emitter has been investigated using pulsed laser excitation. **Figure S5b** displays the time-resolved PL of the emitter, which fits excellently with a single exponential decay function, yielding a lifetime of 5.4 ns, matching well with other SPEs in C-doped h-BN (*26*).

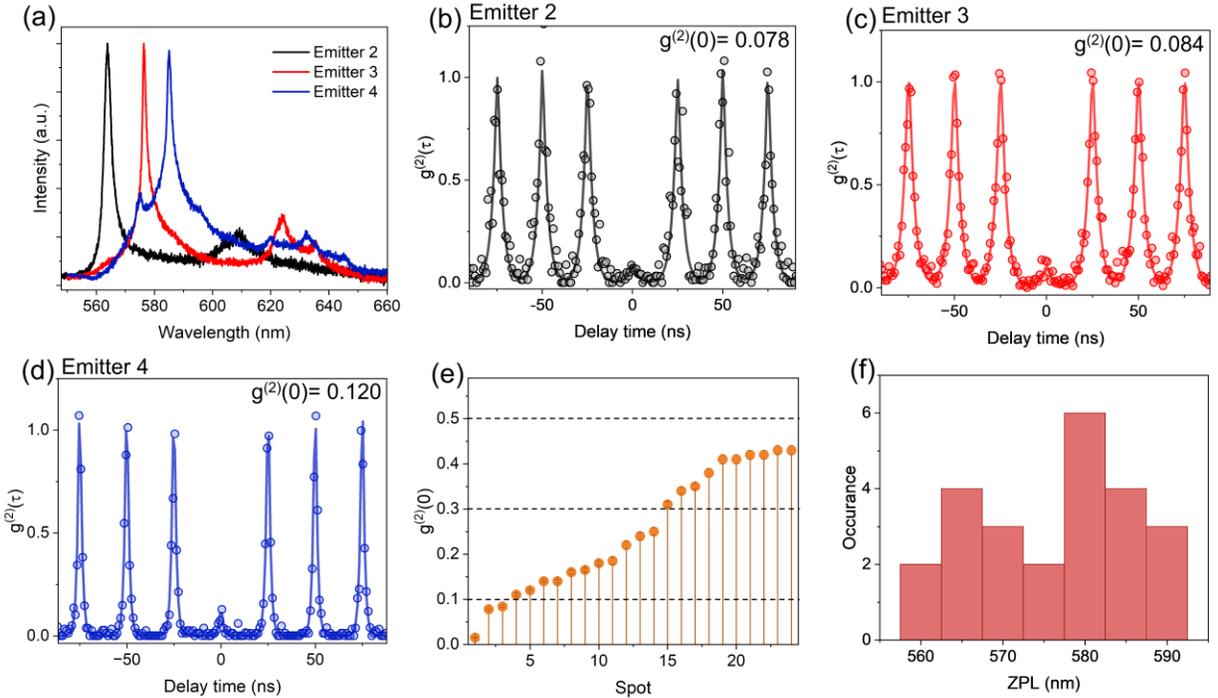

*Figure 3: Optical properties of a few other emitters and distribution of SPEs. (a) Room temperature PL spectra of a few other emitters. Second-order autocorrelation functions, $g^{(2)}(\tau)$ measured for (b) Emitter 2, (c) Emitter 3, and (d) Emitter 4, using 35 μW, 515 nm pulsed laser with a repetition rate of 40 MHz and a pulse width of 96 ps (without background correction). (e) Scatter plot of $g^{(2)}(0)$ for the observed SPEs using 35 μW, pulsed 515 nm laser excitation. (f) Distribution of ZPLs for the observed SPEs.*

In order to investigate the presence of the metastable state in the transition process, the autocorrelation function has been further measured over a longer time scale of up to 0.1 s. As shown in **Figure 2c**, a bunching feature could be seen in this scale, confirming the presence of a shelving state during the optical transition. Upon excitation, the electrons predominantly relax back to the ground state, but small parts relax through the shelving states (*26*). Upon fitting with a multiexponential decay function, the lifetime components (relative weights) related to the metastable state are extracted as 30 μs (94%) and 3.17 ms (6%). This suggests the existence of two metastable states in optical transition. Given the radiative decay lifetime of 5.4 ns, the transition rates involving metastable states are approximately 6,000 and 600,000 times slower, demonstrating predominant relaxation through a radiative channel, marking its high efficiency as a single-photon source.

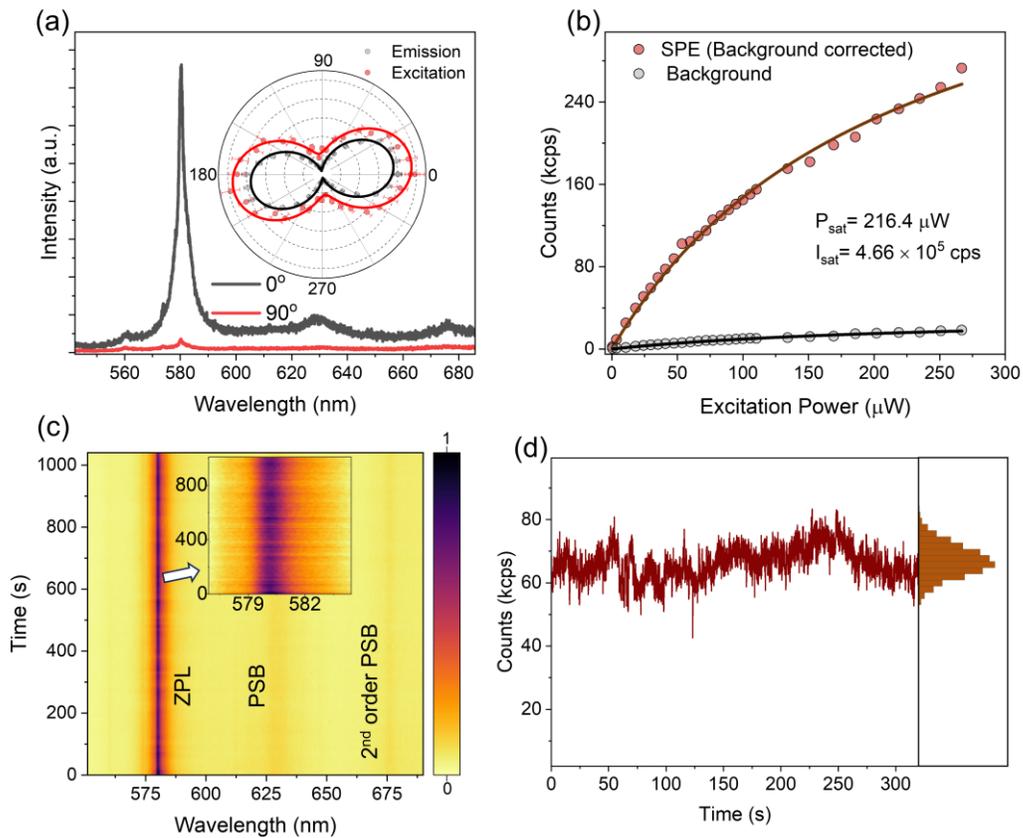

***Figure 4: Polarizability, brightness, and stability of the SPE***. *(a) Polarization-resolved PL spectra of the emitter demonstrate the polarized nature of the emission. Inset showing excitation (red solid circles) and emission (gray solid circles) polarization curves. Solid lines are fits obtained using a cos²(θ) function. (b) Background corrected emission rate as a function of laser (515 nm pulsed excitation) excitation power. The emission rate of the background is also shown for comparison. Solid lines are the fit to the data. (c) Stability test of the PL spectra was performed*

*under an excitation power of 35 μW (532 nm CW) and an integration time of 10 s. The inset shows the zoomed-in portion of the plot, which shows no spectral shift. d) The emission rate (counts/s) of the emitter as a function of time showing superior photostability without blinking measured with 100 ms binning time using 35 μW, pulsed 515 nm laser excitation.*

The PL spectra of several other emitters and their corresponding $g^{(2)}(0)$ values are shown in **Figure 3a-d**, demonstrating the high purity of C-related defects in the thin film, with $g^{(2)}(0)$ values of 0.078, 0.084, and 0.12 for emitter 2, 3, and 4, respectively. To demonstrate the purity of overall emitters in the thin film, the $g^{(2)}(0)$ value of the observed SPEs obtained using the pulsed laser excitation is shown in **Figure 3e**. It has been observed that nearly 12% and 63% of the SPEs exhibit purities exceeding 90% and 70%, respectively, highlighting the effectiveness of generating high-purity SPEs in our C-doped h-BN. As shown in **Figure 3f**, the emitters distributed throughout the films predominantly exhibit emission lines at 575 ± 15 nm, representing an relatively narrower spectral distribution than several other methods (*10*, *15*, *26*). Our findings indicate the preferential formation of a specific defect structure responsible for this emission. The variation of their peak wavelength is attributed to local strain fluctuations in h-BN layers (*39*).

In addition to purity, polarizability, stability, and brightness are also key metrics of SPE. The polarized nature of the emitter was studied by placing an analyzer in the collection path and keeping the excitation polarization fixed. As shown in **Figure 4a,** the PL intensity almost drops to the background level when the analyzer is aligned perpendicular to the polarization direction of the probed emitter. This implies that the emission from ZPL is linearly polarized. To further determine whether the emission center comprises a single dipole or multiple dipoles, both excitation and emission polarization measurements are performed. The inset of **Figure 4a** displays the excitation (red circles) and emission (gray circles) polarization data, along with the corresponding fit obtained using the $\cos^2(\theta-\theta_0)$ function, where $\theta_0$ is the polarization angle (*40*). The polarization angles for excitation (8.2°) and emission (8.6°) were found to be similar. The excitation and emission polarizability are calculated to be 64% and 90%, respectively. These results are characteristics of a single linearly-polarized dipole-like transition (*26*, *41*).

The brightness of the emitter was further evaluated by measuring the emission rate as a function of excitation power. The intensity of the background was also recorded as a function of excitation power to obtain the background corrected intensity of the emitter, as shown in **Figure**

**4b**. The data were then fitted using a first-order saturation model: I = $I_{sat}$ P/(P+$P_{sat}$), where P ($P_{sat}$) and I ($I_{sat}$) are the incident (saturated) power and emission (saturated) rate, respectively. As shown in the figure, the fitted saturation curve yields an emission rate of $I_{sat}$ = 4.66× $10^5$ counts/second (cps), at $P_{sat}$= 216.4 µW, higher than recently reported C and other defect-related SPEs in h-BN (*36, 26, 8*).

Further, we studied the stability of the emitter by recording the consecutive PL spectra. As shown in **Figure 4c**, we do not observe any change in the PL intensity or shift in the ZPL for more than 1,000 seconds of continuous measurement at room temperature, revealing the long-term stability of the emitter. The high stability of few other emitters is also shown in **Figure S4**. Additionally, the emission rate has been monitored for more than 300 seconds to examine the blinking and photobleaching. **Figure 4d** shows the time-dependent emission rate of the emitter with a binning time of 100 ms, demonstrating no blinking or bleaching and thus revealing the emitter's absolute photostability.

*Table 1: Comparison of different characteristics of room temperature SPEs in h-BN obtained through various techniques.*
*MUL: multilayer, FL: few-layer, ML: monolayer

| *Material* | *Treatment* | *ZPL (nm)* | *FWHM (nm)* | *Brightness* | *Lowest $g^{(2)}(0)$ without background correction* | *References* |
|---|---|---|---|---|---|---|
| **Mechanically exfoliated ML flakes** | $^{12}C^+$ implantation | 554-605 | ~9±2 | 200 kcps at 37 µW | 0.23 | (*26*) |
| **Mechanically exfoliated ML flakes** | Ar and H Plasma treatment followed by annealing | 550-750 | ~8±1 | 25 kcps at 0.7 mW | 0.28 | (*8*) |
| **Mechanically exfoliated MUL flakes** | Helium ion irradiation in FIB | 500-700 | ~4±1 | 330 kcps at 2.5 mW | 0.12 | (*36*) |
| **CVD grown ML crystal** | In-situ C doping during growth | ~620 | ~14±1 | - | ~0.25 | (*27*) |
| CVD grown FL film | Thermal annealing in air | ~570 | 4 | - | 0.07 | (20) |
| CVD grown film | Backside boron gettering | 550-600 | ~15±3 | - | 0.2 | (42) |
| PLD grown thin film | In-situ C doping by PLD | 560-590 | 3 | 466 kcps at 216.4 µW | 0.015 | *Our work |

The ultra-high purity, brightness, stability, and DW factor of our SPEs make our sample stand out among the recently reported SPEs in h-BN, as summarized in **Table 1**. We attribute the remarkable optical properties of the emitters to the in-situ doping of h-BN and the low synthesis temperature used in the PLD method, which not only facilitated the preferential formation of a specific type of defect but also preserved the integrity of the crystal structure.

**First-principles calculations**

To understand these experimental observations, we employ first-principles calculations to reveal the structural origins of our exceptional SPEs. Our initial step involved screening C-defect candidates for stability, optical absorption energies that align well with experimental results, and symmetry characteristics, specifically, $C_{2v}$ or $C_s$ symmetry with a mirror plane oriented along either the zigzag or armchair directions of h-BN that could lead to the distinct polarizability observed in our measurements (**Figure 4a**). The defects investigated include vacancies ($V_B$, $V_N$), antisites ($B_N$, $N_B$), C-dopants ($C_B$, $C_N$), and all possible C-dimers formed by a combination of two of these defects ($C_BC_N$, $V_BC_N$, $V_NC_B$, $N_BC_N$, $B_NC_B$), $C_B$–$C_N$, C-dopant donor-acceptor pairs (DAPs) with varying separation distances (2.88 Å and 5.23 Å), C-trimers ($C_BC_NC_B$ or CBC, $C_NC_BC_N$ or CNC) and C-tetramers ($C_BC_NC_BC_N$). The –1, 0, and +1 charge states for each defect were also systematically evaluated following an initial screening for stable charge transitions using the Perdew-Burke-Ernzerhof (PBE) (43) and Heyd-Scuseria-Ernzerhof (HSE06) (44) functionals (**Figure S6**). To identify the defects most likely to explain single photon emission, we applied HSE06 combined with the independent particle approximation (IPA) to refine our initial candidate selection to those with polarized absorption within 0.75 eV of our experimental ZPL value ~2.10 eV (a range of 1.35 eV to 2.85 eV). We also excluded candidates with an additional lower energy peak with stronger absorption than our potential SPEs. This approach is chosen for initial screening due to the previously established high fidelity of HSE06 and IPA in predicting optical band gaps and defect absorption energies (e.g., within 10% mean absolute error), even without considering the excitonic interactions typically captured in many-body perturbation theory (*45–47*). Although absorption alone is not a direct predictor of single photon emission, the experimentally measured S factor is relatively low (0.79), indicating weak electron-phonon coupling in our SPEs. This suggests a small reorganization energy and a modest Franck-Condon (FC) shift. Consequently, it is unlikely for a defect with an absorption feature exceeding 2.85 eV to simultaneously support emission around 2.10 eV while maintaining weak electron-phonon coupling. All final candidates are presented in **Figure 5a**, while discarded structures appear in **Figure S7-9**.

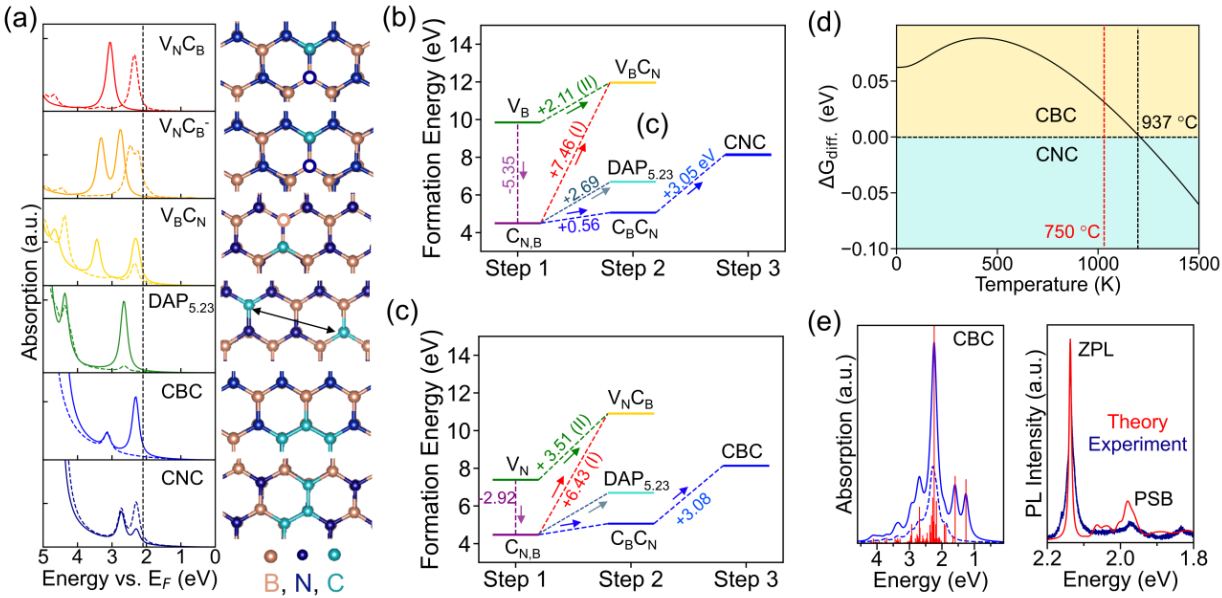

*Figure 5: First Principals Calculations on Defects for SPEs. (a) HSE06-IPA absorption calculations for defects with polarized absorption peaks within 0.75 eV of the observed experimental ZPL energy. Each defect is visually represented next to its corresponding theoretical spectra, and the ZPL energy from the experiment is marked with a dashed black line. (b,c) HSE06 calculated formation enthalpies for the two proposed formation mechanisms: the dopant-dopant complex pathway and the dopant-vacancy complex pathway. The pathways are compared using (b) boron vacancies and (c) nitrogen vacancies, respectively. For PBE calculations, see **Figure***

***S10***. *(d) Gibbs free energy difference between the two C trimers (CBC and CNC) as a function of temperature where positive values indicate a more stable CBC phase and negative values a more stable CNC phase. The black and red dashed lines signify crossover and our growth temperature, respectively. (e) $G_0W_0$-BSE calculated absorption spectra of the CBC defect (left panel). The oscillator strengths are shown as red vertical lines. The calculated PL spectra (right panel) for the defect (red) are compared with the experiment (dark blue). For all spectral plots (HSE06-IPA) in (a) and $G_0W_0$-BSE in (e), polarization along the x-axis is shown as solid-colored lines and y-axis dashed-colored lines.*

Our calculations identify six candidate emitters: CNC, CBC, DAP$_{5.23}$, V$_B$C$_N$, V$_N$C$_B^-$, and V$_N$C$_B$. Each of these candidate structures has been proposed separately in independent studies (*25, 26, 48–51*). To evaluate these candidates further, we calculated the ZPL energy for each structure using constrained DFT at the HSE06 level (ΔSCF method) and confirmed that each has a radiative transition energy within 0.5 eV of our experiments (see **Table S1** for ZPL energies), which supports prior claims on their suitability for SPEs (*25, 26, 49–51*). These defects can be broadly separated into three categories: vacancy-dopant complexes (V$_B$C$_N$, V$_N$C$_B$, and V$_N$C$_B^-$), dopant-dopant complexes (CBC and CNC), and range-separated dopant DAPs (DAP$_{5.23}$). To evaluate the likelihood of formation for each class of defect complexes, we compare the thermodynamic formation enthalpy (ΔH) calculated for the two proposed formation mechanisms, shown in **Figure 5b-c** at the HSE06 level and in **Figure S10** at the PBE level.

In the first proposed pathway, C initially replaces a lattice ion to form a substitutional dopant (C$_B$ or C$_N$) at time zero (step 1 in **Figure 5b**), while in the second proposed pathway, a vacancy is introduced (V$_B$ in **Figure 5b**, V$_N$ in **Figure 5c**). We then assess the favorability of forming vacancy-dopant complexes, dopant-dopant complexes or dimers, and DAPs starting from time zero of each pathway. For the first pathway, the energy required to form a C-complex is relatively small for dimers (C$_B$C$_N$, +0.56 eV). DAPs require moderately higher energy (+2.69 eV, approximately 4.5 times larger than ΔH$_{dimer}$), but the energy required for expelling a lattice N atom to form a vacancy-dopant dimer is over an order of magnitude larger (e.g., +7.46 eV for V$_B$C$_N$ in **Figure 5b**) than dimer formation. The energy gain for C doping a lattice N atom directly adjacent to an existing B vacancy to form V$_B$C$_N$ (+2.11 eV, step 2 in pathway II in **Figure 5b**) is substantially lower than the N expulsion step in pathway I. However, a competing step also present in the second

pathway – C-dopants directly passivating high-energy vacancies ($V_B + C \rightarrow C_B$ in **Figure 5b**) – yields a large reduction in energy (-5.35 eV). Similar findings are observed for $V_NC_B$ (**Figure 5c**). Collectively, these results imply that C-vacancy complexes are unlikely to form in h-BN, and it is far more energetically favorable to passivate an existing vacancy than to form a vacancy-C complex. After C-doping occurs, it is energetically unfavorable to expel a lattice B or N atom adjacent to the dopant. Therefore, we restrict all further calculations and analysis to C-C complexes and DAPs that more directly match our experimental conditions.

From **Figure 5b-c**, C-C dimers ($C_BC_N$) appear highly stable compared to other defect pairs, their HSE06-IPA absorption energies are too high to account for our experimental results (**Figure S9**). Indeed, C-dimers have been proposed as a potential ~ 4 eV emitter in a separate study (*52*). Hence, single photon emission likely arises either from the formation of less energetically favorable DAPs (e.g., caused by kinetic trapping driven by experimental factors not captured in DFT) or from trimers (CBC or CNC) that form as a secondary step following dimer formation. Trimer formation (step 3 in **Figure 5b-c**) requires an additional 3.05/3.08 eV, which is still lower than the formation energy of the competing reaction where a dimer forms with an isolated C-dopant (e.g., at the HSE06 level $C_BC_N + C_B$ is ~ 1.4 eV larger in energy than CBC). Interestingly, the two trimers exhibit strikingly similar formation enthalpies (differing by ~60 meV in PBE or ~30 meV in HSE06) at 0 K, suggesting the preferred trimer phase is highly sensitive to temperature in stoichiometric h-BN.

To explore this possibility, we employed density functional perturbation theory (DFPT) calculations at the PBE level to determine the temperature-dependent Gibbs free energy (ΔG) for each structure and plot the stability of different defects (difference in ΔG between the two phases) at constant C concentration (CBC vs. CNC trimers in **Figure 5d** and C dimers vs. DAPs in **Figure S11**). We found that the CBC trimer is favored at lower temperatures (below ~940 °C), while CNC is favored at higher temperatures. By contrast, DAPs have a much higher formation enthalpy than dimers and exhibit a weaker ΔG temperature dependence (**Figure S11**). It is worth noting that the relative entropic stabilization of DAPs at elevated temperatures would likely be much higher than our calculations predict if we compared dimer formation against all possible DAP configurations. However, in this analysis, we only considered DAP configurations that satisfy our previously outlined criteria for HSE06-predicted optical transitions. These results strongly indicate that dimer formation in h-BN is highly robust despite lacking optical transitions in the experimentally observed range. While we cannot entirely rule out DAP formation due to experimental kinetic factors not captured by DFT calculations, the strong thermodynamic preference for dimers over DAPs suggests that, if DAP formation does occur, it likely does so alongside and at lower concentrations than dimers, which potentially facilitate subsequent trimer formation. The similarity in formation energies for both trimers **(Figure 5b-c)** likely indicates that both form under many experimental conditions. This result is consistent with recent microscopy studies showing a strong preference for C clustering into dimers and trimers (*53, 54*). In our study, the relatively lower synthesis temperature (750 °C) compared with previous studies (ranging from 1045 °C to 1350 °C ) suggests that our PLD method favors CBC creation, which we expect to exhibit distinct optical properties from CNC and DAPs (*25, 27*).

Further, we recalculated the optical absorption spectra (**Figure 5e** for CBC and **Figure S12** for CNC and $DAP_{5.23}$) and determined the excitonic properties (**Table 2**) of the final three candidates (CBC, CNC, and $DAP_{5.23}$) with $G_0W_0$-BSE (see **Figure S13** for the orbital diagram and projected wavefunctions). To the best of our knowledge, these are the first calculations performed for bulk C-doped h-BN at the $G_0W_0$-BSE level, which are typically done at the monolayer level where bulk dielectric screening and interlayer coupling effects play a major role in modifying optical properties (*26*). The optical transition energy for the absorption peak with the highest oscillator strength within the 0.1– 4.5 eV range is presented in **Table 2** as $E_{abs}$. Additionally, we employ constrained DFT at the PBE level to calculate the FC shift, which is subtracted from $E_{abs}$ to determine the ZPL energy. The PL spectra for the three defects were also calculated (right panel of **Figure 5e** for CBC and **Figure S14** for CNC and $DAP_{5.23}$), which exhibit excellent agreement with the experimental data. Subsequently, we compared several additional optical properties to identify the most promising SPE candidate (**Table 2**). These properties include the square modulus of the exciton dipole moment ( $|\mu^2_{(e-h)}|$), radiative lifetime ($\tau_R$) calculated using Fermi's golden rule (*55, 56*),  the nuclear coordinate change or excited-state reorganization ($\Delta Q$) between initial and final states from constrained DFT, electron-phonon coupling effects including the S as well as DW factor and the LO phonon energy ($E_{LO}$). Full spectral function plots used to derive the S factor can be found in **Figure S15**.

*Table 2: Comparison of the calculated optical properties for the defect emission.*

| Defect | $E_{abs}$ (eV) | FC (eV) | ZPL (eV) | $E_{LO}$ (meV) | $\mu^2_{e\text{-}h}$ (a.u.) | $\Delta Q$ (amu$^{1/2}$ Å) | $\tau_r$ (ns) | S factor | DW factor |
|--------|------|------|-------|------|------|------|------|------|------|
| **DAP** | 2.60 | 0.52 | 2.08 | 172 | 4.86 | 0.46 | 11.0 | 3.17 | 0.04 |
| **CNC** | 2.03 | 0.19 | 1.84 | 159 | 4.84 | 0.31 | 23.0 | 1.36 | 0.26 |
| **CBC** | 2.24 | 0.13 | 2.11 | 158 | 6.33 | 0.26 | 13.2 | 0.99 | 0.37 |
| Exp. | - | - | 2.136 | 166 | - | - | 5.4 | 0.79 | 0.45 |

As seen from **Table 2**, while each defect exhibits predicted ZPL energies that align closely with the experiment, CBC stands out as the most promising SPE candidate due to its large transition dipole moment, weak electron-phonon coupling, and relatively shorter radiative lifetime. The predicted emission peak energy, radiative lifetime, PSB, S, and DW factor are all in strong agreement with experimental data. Although DAPs demonstrate a slightly faster radiative lifetime, their large excited-state reorganization and FC shift lead to stronger electron-phonon coupling. This results in a higher S factor, broader linewidths, lower DW factor, and, correspondingly, reduced single photon purity. In addition, DAPs are not energetically favorable, as discussed above. CNC defects exhibit optical properties close to those of CBC and are also highly probable but with typical higher-temperature synthesis protocols used in other studies (*27*). Additionally, CNC displays slightly lower performance in key optical metrics, including weaker oscillator strength, a reduced transition dipole moment, a longer radiative lifetime, and higher ΔQ, which increases electron-phonon coupling. Together with our thermodynamic analyses, these findings suggest that the remarkably high brightness, large DW factor, and high stability of our SPEs in our experiments can be attributed to the low-temperature synthesis method in PLD, which favors trimer formation with CBC as the dominant structure.

## DISCUSSION

In conclusion, we demonstrated a PLD method for creating C-related defects in a centimeter-scale h-BN thin film. These defects function as SPEs that exhibit excellent single photon emission properties, including record high purity, luminescence stability, brightness, and polarizability, which are critically important for practical applications of SPEs. The emitters also display prominent emission lines at (575 ± 15) nm, indicating a preference for forming a specific defect structure responsible for this emission. Moreover, employing comprehensive first-principles calculations, we identified trimers ($C_BC_NC_B$) as suitable defects to explain the observed results. This defect exhibits superior optical properties compared to most others and is energetically favored in our experiment due to the low synthesis temperature achieved with our PLD method. The high quality of the SPEs developed in this work makes them highly promising for various quantum applications. Moreover, our synthesis approach offers an efficient and reliable method for producing quantum emitters. Our calculations and analysis further provide essential insights and guidelines for optimizing quantum optical emissions and advancing quantum material synthesis in the future.

## MATERIALS AND METHODS

### PLD growth of C-doped h-BN thin film

C-doped h-BN films were grown by slightly modifying the previously reported PLD process, utilizing a KrF excimer laser with a wavelength of 248 nm and a pulse width of 25 ns (*22*, *57*). Films were grown by using the following deposition conditions: growth temperature ~750 °C, laser fluency ~ 2 J/cm$^2$ (laser energy ~220 mJ), repetition rate 5 Hz, target to substrate distance ~ 50 mm, and in the presence of partial high-purity (5N) nitrogen background pressure of 100 mTorr. We used commercially available high-purity (99.9% metal basis) one-inch diameter polycrystalline 1 at% C-doped h-BN target (American Element) for the laser ablation. Meanwhile, an undoped h-BN target was used for pristine h-BN thin film. For deposition, we used highly insulating double-sided polished *c*-Al$_2$O$_3$ (0001) sapphire substrates (University Wafer). Prior to growth, we prepared the atomically flat surface of the substrates by annealing them at 1200 °C for approximately 2 hrs. Before mounting inside the PLD chamber, the substrates were sonicated in acetone for 15 minutes and then dried on a hot plate at 100 °C. In-situ, pre-annealing of the substrates was also conducted at the growth temperature and pressure for about 15 minutes. We used 2000 laser shots to achieve a film thickness of ~32 nm. After growth, films were cooled down at a rate of ~20 °C/min.

## Structural characterization of h-BN

XPS was performed using PHI Quantera SXM scanning X-ray microprobe equipped with a 1486.6 eV monochromatic Al Kα X-ray source. High-resolution core-level elemental scans for B 1s, N 1s, and C 1s were recorded at a pass energy of 26 eV. Park NX20 AFM was used to obtain surface topography, operating in tapping mode using Al-coated Multi75Al cantilevers. Raman spectroscopy was carried out using a Renishaw inVia confocal microscope with a 532 nm CW laser with a spot size of 1 μm as the excitation source. The depth profile analysis of elements was performed using time-of-flight secondary-ion mass spectrometry (ToF-SIMS, ION-TOF GmbH, Münster, Germany) combined with an in-situ scanning probe microscope (NanoScan, Switzerland) and a $Cs^+$ ion beam for sputtering.

## Photophysical measurements on thin film

All the PL measurements were carried out using Renishaw inVia confocal microscope equipped with a CCD camera. The spectra were collected using a 100× high NA (0.9) objective. Emission and excitation polarizations were measured by placing a broadband analyzer and a half-wave plate in the optical collection and excitation path, respectively. Second-order autocorrelation measurements and related PL raster scanning were carried out using a PicoQuant MicroTime 100 upright time-resolved photoluminescence microscope. The excitation source was a pulsed diode laser emitting at 515 nm. The laser was used with a repetition frequency of 40 MHz and 96 ps pulse width. The laser was directed to a high numerical aperture (100×, NA = 0.9) objective lens. Imaging and positioning the laser spots were carried out via objective scanning using a 3D piezo scanner. The detection light was filtered with a 532 nm dichroic mirror followed by a long pass 550 nm filter before coupling into a graded-index multimode fiber (0.22 NA) with a fiber aperture of 62.5 μm acting as a confocal pinhole. A flip mirror directed the emission either to a spectrometer (FluoTime 300, PicoQuant) or to two avalanche photodiodes (Excelitas Technologies) in an HBT configuration for the correlation measurements. A 50:50 beam splitter was used to divide the incoming photon, followed by a bandpass filter placed before each of the detectors. The lifetime measurements were carried out in the FluoTime 300 spectrometer equipped with a PMA-Hybrid 50 detector.

## Electronic Structure Calculations

DFT calculations employed the plane-wave projector augmented (PAW) method as implemented in the Vienna ab initio simulation package (VASP) (*58*). Bulk h-BN lattice parameters were determined by separately optimizing the h-BN unit cell with PBE (*43*) and HSE06 (*44*). The PBE optimized lattice formed the basis for all supercell calculations at the PBE or $G_0W_0$-BSE level, and the HSE06 lattice formed the basis for all HSE06 supercell calculations. The stability of each reported defect and its charge states were initially screened by converting the optimized lattices to a 3×3×2 supercell (72 atoms), introducing a defect, and relaxing the structure. The reported stabilities were calculated using equation 1 as implemented in the Spinney package (*59*).

$$E_{form} = E_{form}(\text{def}) - E_{total}(no\ \text{def}) - \sum_\alpha n_\alpha \mu_\alpha + qE_F + E_{corr} \quad \text{......................(1)}$$

where $E_{tot}$ (def) and $E_{tot}$ (no def) corresponds to the change in total energy between defective and defect-free supercells, $n_\alpha$ represents the number of atoms of species alpha added (+1) or removed (-1), $\mu_\alpha$ denotes the chemical potential of alpha, q the defect charge, $E_F$ the fermi level, and $E_{corr}$ the FNV correction for finite-size effects in charged defect calculations.(*60*) Linear response theory (independent orbital approximation/single particle level) was used to simulate all HSE06-IPA absorption spectra in accordance with the procedure described in the literatures (*46*). After initial screening of all solo, dimer, trimer, and tetramer defects by HSE06-IPA and both PBE and HSE06 for formation energy, a larger cell (5×5×1) was used for DAP calculations to allow for larger lateral separation between defects. C-trimer and dimer absorption spectra and formation energies were recalculated for the same supercell size and did not shift by more than 0.02 eV for formation energy or 0.05 eV in the HSE06-IPA peak. All final theory-predicted properties for DAPs, C-dimer, and C-trimers (including later described $G_0W_0$-BSE calculations) utilized the larger 5×5×1 supercell. The Gibbs free energy for the two C-trimers, C-dimers, and DAP$_{5.23}$ were determined using equation 2:

$$\Delta G = U_{struc} + U_{vib} - T\Delta S \quad \text{......................(2)}$$

where $U_{struc}$ is the enthalpic formation energy calculated with VASP and equation 1, $U_{vib}$ is the vibrational formation energy of each defect calculated via DFPT within the Phonopy code (*61*), and $\Delta S$ is the sum of the configurational entropy as calculated by Stirling's approximation and the vibrational entropy calculated in Phonopy. As described in the main text, once the Gibbs free energy is calculated for each defect, the difference between the two trimers is plotted in **Figure 5d**, and between the dimer versus DAP$_{5.23}$ in **Figure S11**. For predicting optical transitions, including excitonic effects, we calculated the G$_0$W$_0$-BSE absorption spectra using the PBE-optimized geometry and orbitals. The FC shift and $\Delta Q$ were calculated using constrained DFT at the PBE level, and the FC shift from PBE was subtracted from the calculated G$_0$W$_0$-BSE E$_{abs}$ energy to determine the G$_0$W$_0$-BSE ZPL energy. For the HSE06 predicted ZPL energies, the $\Delta$SCF method was used. The PL line shapes were simulated using the PyPhotonics package using the PBE ground and excited state geometries and phonon band structure as determined by the finite difference method in the Phonopy code (*62*). All DFT calculations employed a plane-wave cutoff energy of 800 eV. The h-BN unit-cell geometry and cell-shape optimizations used a 10×10×10 Γ-centered mesh for both PBE and HSE06 calculations. However, the K-mesh size was reduced to 4×4×1 (PBE) and 2×2×1 (HSE06) for relaxing atomic positions in the defect supercells, and determining formation energies and stable charge-transitions. HSE06-IPA calculations used the charge density from the 2×2×1 mesh single point calculations but with nine additional zero-weighted KPOINTS along the Γ-M-K high symmetry pathway (3 points between each high symmetry KPOINT). All PBE and HSE06 geometries and atom positions were relaxed until residual forces were below 0.01 eV/Å. G$_0$W$_0$-BSE calculations were performed using a Γ-centered 3×3×1 KPOINT mesh for the PBE 5×5×1 supercell ground-state geometries starting from the pre-converged PBE wave functions.

# Acknowledgments

## Funding

A.C and S.H acknowledge the funding support of the National Science Foundation (ECCS-2230400, ECCS-1943895, and ECCS-2246564), Welch Foundation (C-2144), and the Air Force Office of Scientific Research (FA9550-22-1-0408). A.B. and P. M. A. acknowledge the funding supported by the U.S. Air Force Office of Scientific Research and Clarkson Aerospace Corp. (FA9550-24-1-0004). I.A. acknowledges financial support from the Australian Research Council (CE200100010, FT220100053) and the Office of Naval Research Global (N62909-22-1-2028).

## Author Contributions

A.C. and A.B. conceived the idea. A.B. performed material synthesis and helped in material characterization. A.C. designed and performed all the measurements and analyzed the data with the support of S.H. T.T. performed the TOF-SIMS measurements and helped in the analysis. I.A. helped in the analysis of data and provided helpful directions. A.S.F. and B.J.S. performed first principles calculations. A.C. wrote the manuscripts with support from A.B., A.S.F., and S.H. All authors contributed to the revision of the manuscript. S.H. supervised the project.

## Competing Interest

All the authors declare no competing interest.

## Data and materials availability

All data needed to evaluate the conclusions in the paper are present in the paper and/or the Supplementary Materials.

## Table of Contents for Supplementary Materials

Figure S1 to S15, Table S1

# Supplementary Materials for

**Room-Temperature High-Purity Single Photon Emission from Carbon-Doped Boron Nitride Thin Films**

Arka Chatterjee *et al.*

*Corresponding Author. E-mail: shengxi.huang@rice.edu*

**This PDF file includes:**

Figures S1 to S15

Table S1

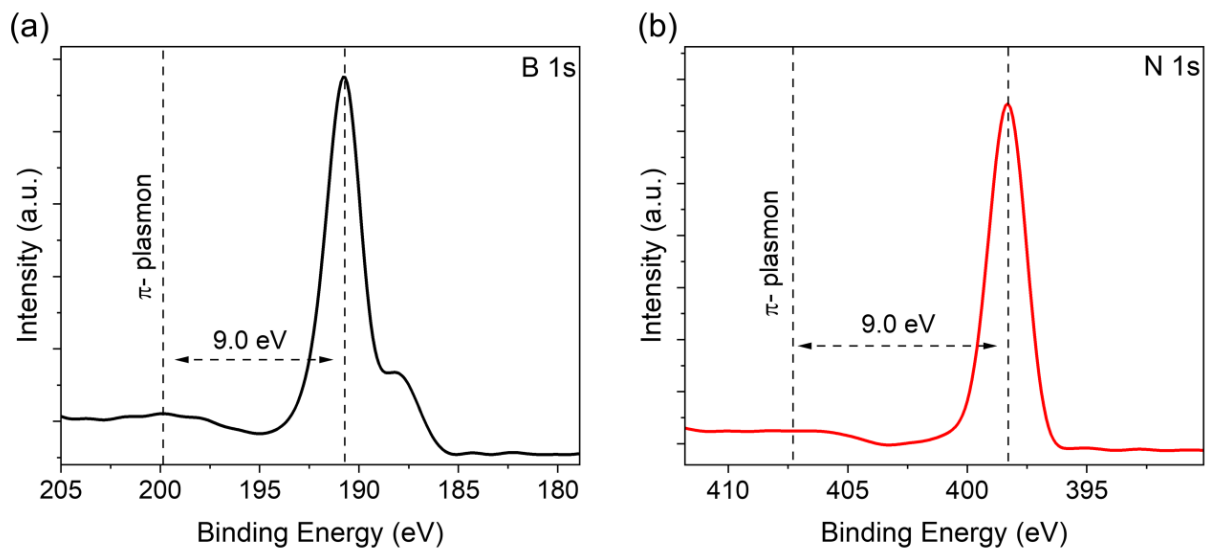

*Figure S1*: ***XPS analysis of B 1s and N 1s spectra.*** *XPS spectra of B 1s and N 1s in a wider energy scale showing the π-plasmon peaks.*

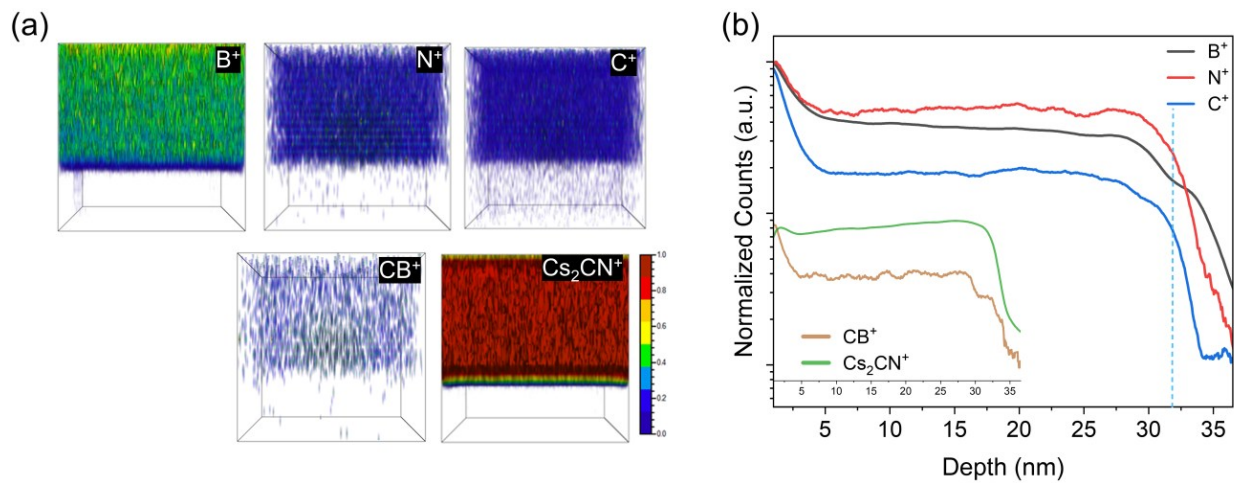

*Figure S2*: **Depth profile analysis from TOF-SIMS measurements.** *(a) Transversal view of 3D intensity plot (150 μm × 150 μm × 60 nm) for $B^+$, $N^+$, $C^+$, $CB^+$, and $Cs_2CN^+$. (b) Depth profile of all elements.*

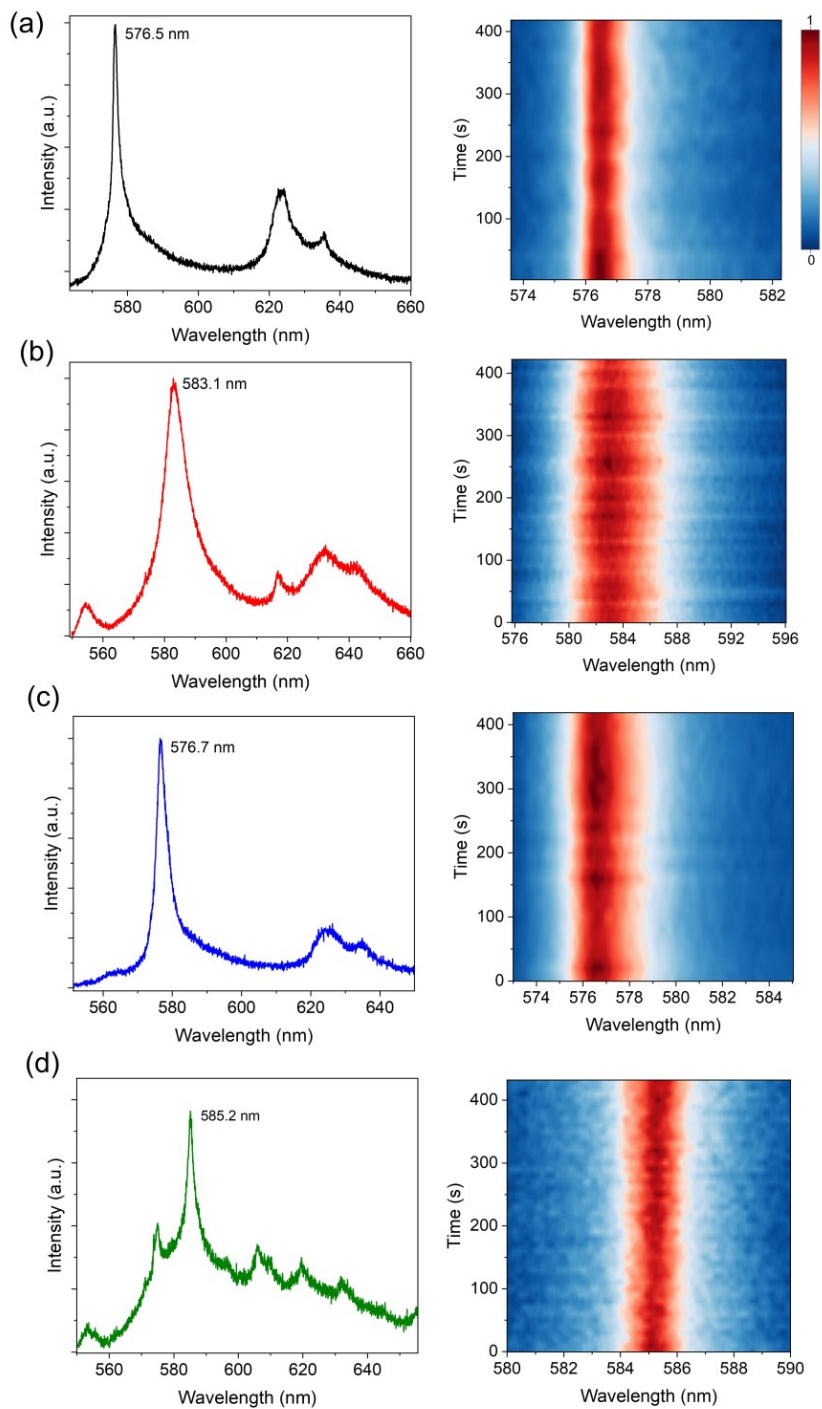

*Figure S3: Uniformity and stability of the emitters.* *(a)-(d) The PL spectra of other four other SPEs (Left panel). Stability of the ZPLs for the SPEs (Right panel).*

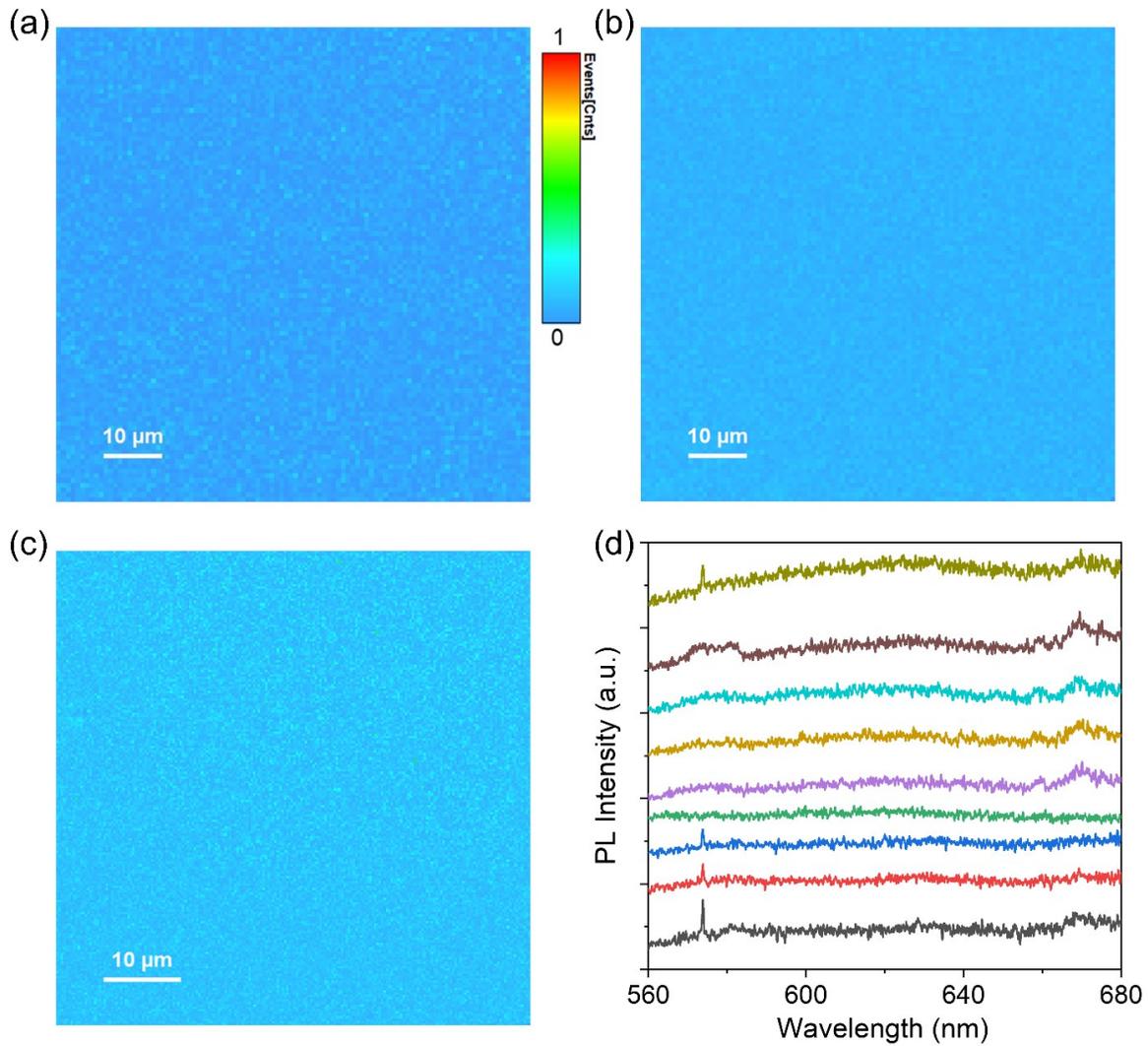

***Figure S4****: **Spectroscopic studies on pristine h-BN thin film***. *(a)-(c) The confocal PL mapping of pristine h-BN sample at three different regions demonstrating the absence of any SPEs. (d) PL spectra acquired at multiple locations in the mapped regions of pristine h-BN do not exhibit characteristics indicative of single-photon emission.*

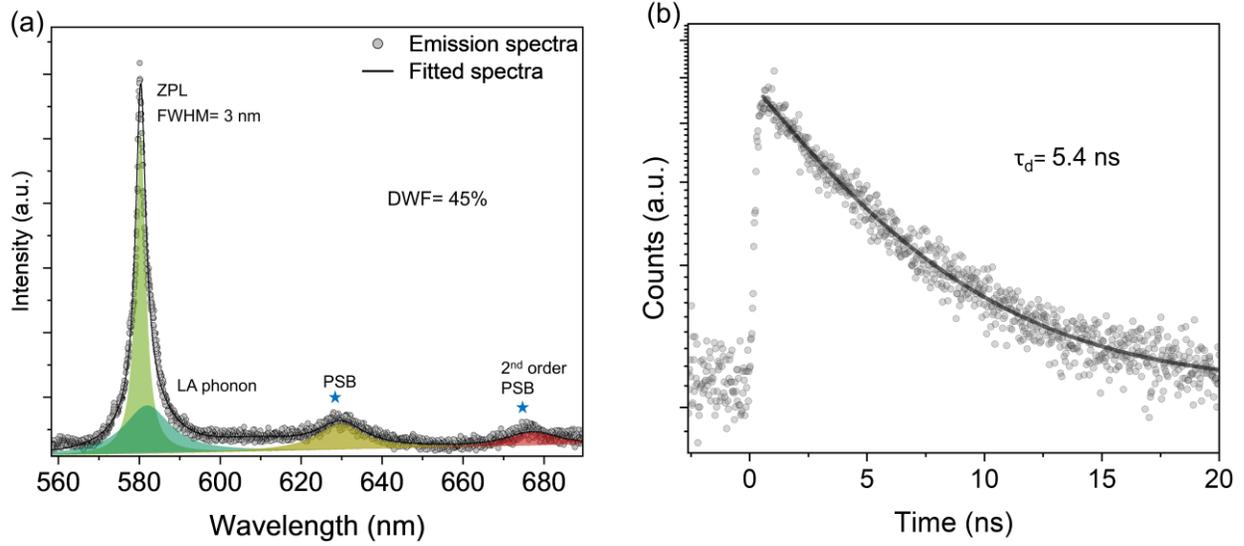

*Figure S5: **Fitted PL spectra and TRPL decay of the SPE.** (a) The PL spectrum fitted to obtain the individual weightage of ZPL, LA, and LO phonons. (b) Time-resolved PL of the emitter exhibiting an excited state lifetime of 5.4 ns. A 35 µW, 515 nm pulsed laser with a repetition rate of 40 MHz and a pulse width of 96 ps was used as the excitation source. The solid lines are fits obtained using single exponential decay functions.*

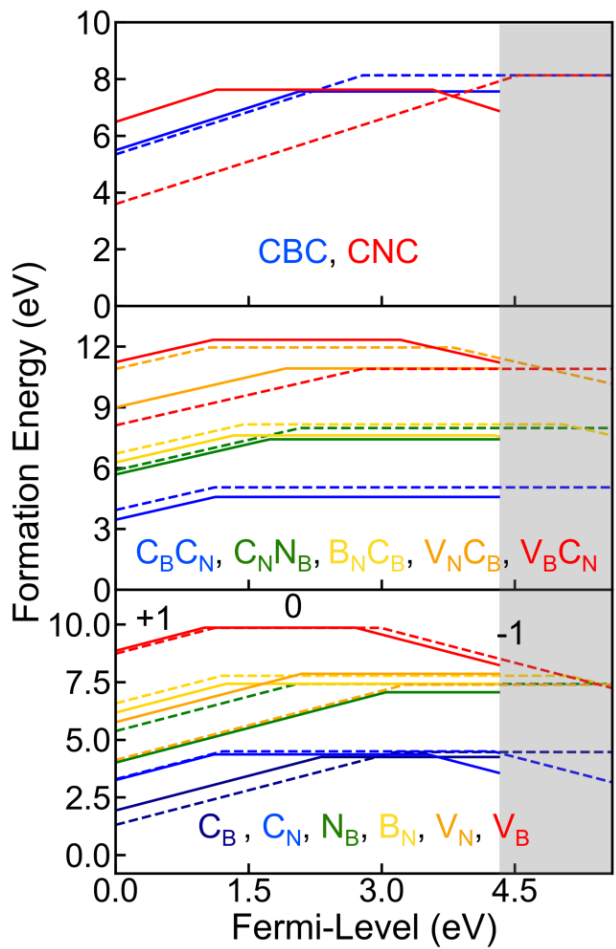

*Figure S6*: ***Charge transition energies and Fermi-level analysis for screened defects in h-BN.*** *Formation energies of all screened defects and their corresponding charge-transitions. The positive slopes represent the +1 charge state, flat lines indicate the 0 charge state, and the negative slope corresponds to the –1 charge state. Dashed lines represent calculations using HSE06, whereas solid lines represent the PBE calculations. The Fermi-level is in respect to the h-BN valence band (e.g., a Fermi-level of 0 eV is exactly at the top of the h-BN valence band). The HSE06 band gap for h-BN is closer to the experiment (5.5 eV), and the Fermi-level range covers the full HSE06 band gap (e.g., Fermi-level of 5.5 eV corresponds with the bottom of the HSE06 predicted h-BN conduction band). The PBE band gap is expected to be smaller as PBE is known to underestimate experimental band gaps. Therefore, the Fermi-level range for PBE calculations was limited to a smaller window. The grey shading represents the difference between PBE and HSE06 h-BN band gaps.*

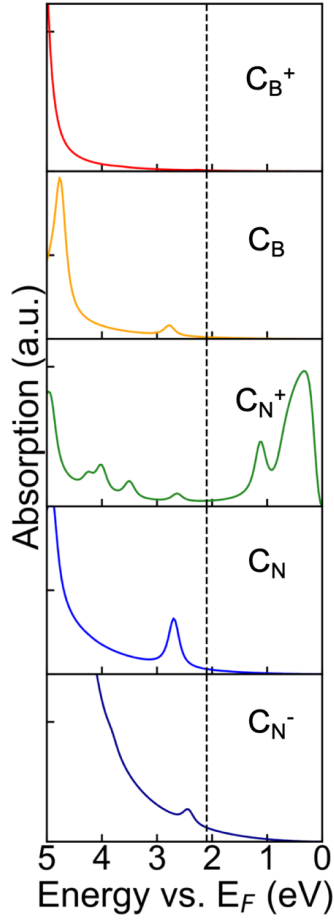

*Figure S7*: ***Theoretical absorption energies and polarization behavior of solo C-defects in h-BN.*** *HSE06-IPA predicted absorption energies for each solo C-defect with stable charge transitions. Experimental predicted ZPL is shown as a dashed vertical line. Similar to the main text, polarization along the x-axis is shown as solid colored lines and the y-axis as dashed colored lines. For cases where there is no polarization (x and y absorption is equal), the two spectra perfectly overlap. This implies isolated C-defects do not show any polarization behavior, indicating inconsistent optical properties with the present SPEs.*

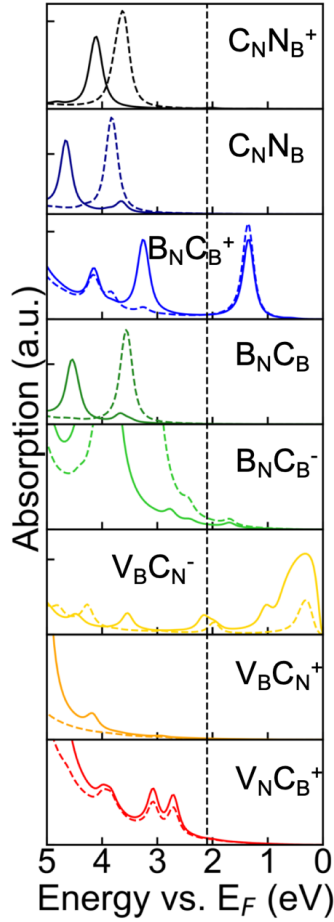

*Figure S8*: ***Theoretical absorption energies and polarization behavior of combination of defects in h-BN.*** *HSE06-IPA predicted absorption energies for vacancy-C dopant and antisite-C dopant type defects with stable charge transitions. Experimental predicted ZPL is shown as a dashed vertical line. Similar to the main text, polarization along the x-axis is shown as solid colored lines and the y-axis as dashed colored lines. For cases where defect structures met the screening criteria outlined in the main text for SPE, their spectra are shown in **Figure 5a** of the main text.*

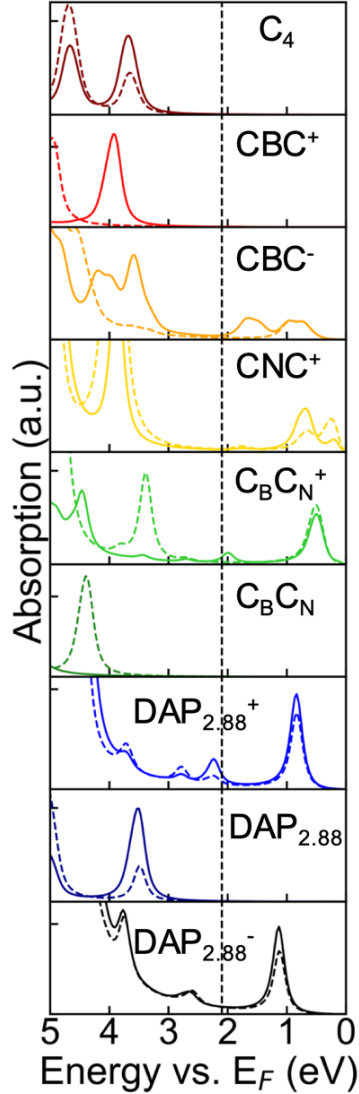

*Figure S9: **Theoretical absorption energies and polarization behavior of combination of C-dopant defects in h-BN.** HSE06-IPA predicted absorption energies for dimer, trimer, tetramer, and DAP defects with stable charge transitions. Experimental predicted ZPL is shown as a dashed vertical line. Similar to the main text, polarization along the x-axis is shown as solid colored lines and the y-axis as dashed colored lines. The tetramer structure is referred to as C4 and is the trans isomer. For cases where defect structures met the screening criteria outlined in the main text for SPE, their spectra are shown in **Figure 5a** of the main text.*

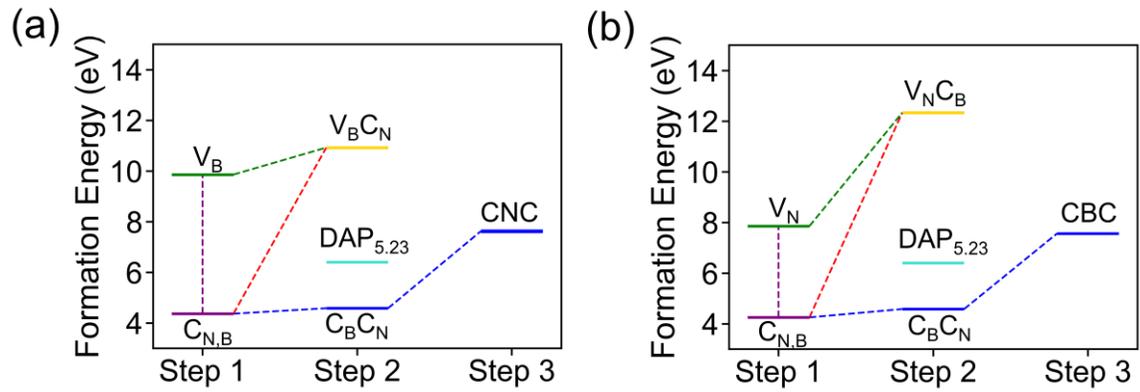

*Figure S10*: **PBE calculated formation enthalpies.** *PBE calculated formation enthalpies for the two proposed formation mechanisms: the dopant-dopant complex pathway and the dopant-vacancy complex pathway. The pathways are compared using (a) boron vacancies and (b) nitrogen vacancies.*

*Table S1:* *HSE06 ΔSCF ZPL energies.*

| Defect | ZPL (eV) |
| --- | --- |
| CBC | 1.63 *(49)* |
| CNC | 1.65 *(49)* |
| DAP$_{5.23}$ | 2.15 *(26, 51)* |
| V$_B$C$_N$ | 161 *(25, 51)* |
| V$_N$C$_B{}^-$ | 1.82, 1.51 *(50)* |
| V$_N$C$_B$ | 1.72 5 *(50)* |

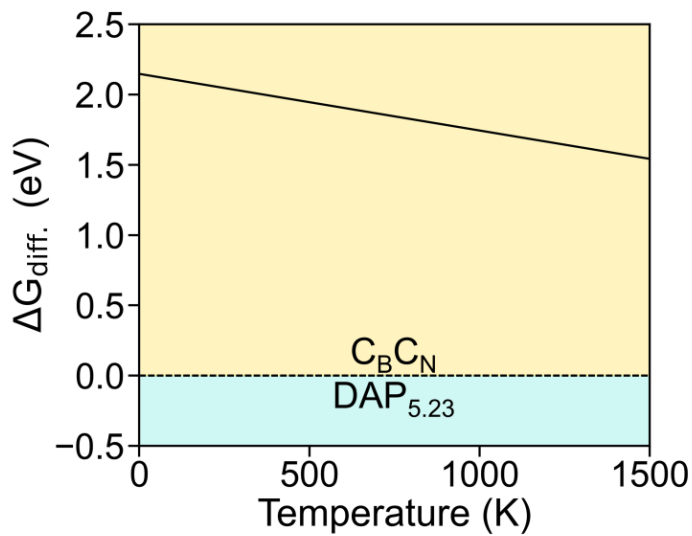

*Figure S11: Gibbs free energy difference between the C dimers ($C_BC_N$) and $DAP_{5.23}$ as a function of temperature.* *Gibbs free energy difference between the C dimers ($C_BC_N$) and $DAP_{5.23}$ as a function of temperature where positive values indicate a more stable dimer phase and negative values a more stable DAP phase. The consistently positive value indicates the dimer phase is more stable across the entire temperature range. However, the decreasing magnitude indicates the relative stability is smaller at higher temperatures.*

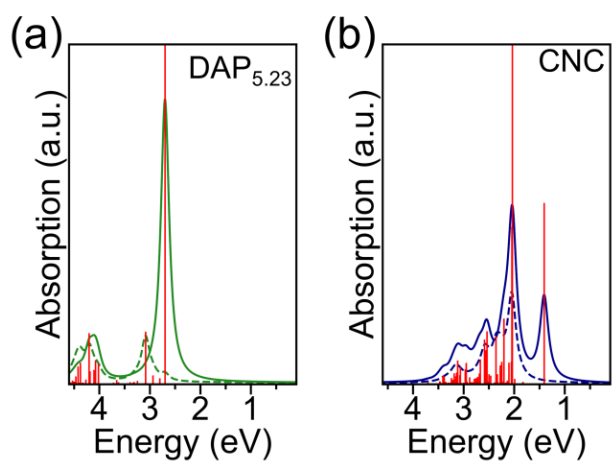

*Figure S12: Calculated absorption spectra of DAP$_{5.23}$ and CNC.* $G_0W_0$-BSE *calculated absorption spectra of (a) DAP$_{5.23}$ and (b) CNC. The $G_0W_0$-BSE absorption spectra are shown using the same colors as their corresponding defect in* **Figure 5a**, *and the oscillator strengths are shown as red vertical lines.*

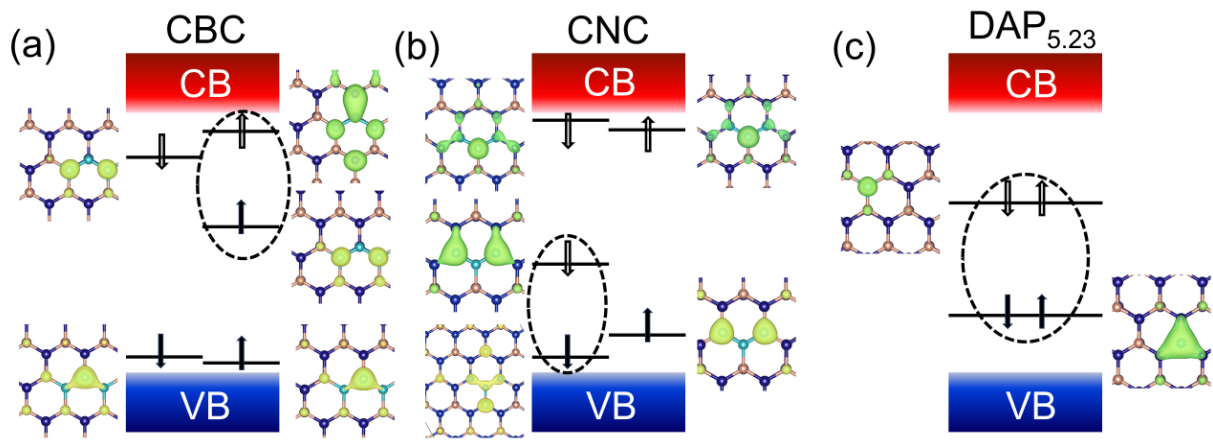

*Figure S13: Orbital diagram for CBC, CNC, and DAP$_{5.23}$.* Orbital diagram for $G_0W_0$-BSE predicted optical transitions for (a) CBC, (b) CNC, and (c) DAP$_{5.23}$. For CBC and CNC, the wavefunction for the marked state is shown on the left for the spin down channel and on the right for the spin up channel. For DAP$_{5.23}$ the predicted orbital diagram is closed-shell. In each case, states below the fermi-level (occupied) are shown as fully black arrows, and states above the fermi-level (unoccupied) are shown as arrows with black outlines and grey infilling. The optical transitions with the strongest oscillator strengths discussed in the main text are marked with a dashed black circle. The VB and CB for each structure are the HOMO and LUMO for h-BN, respectively.

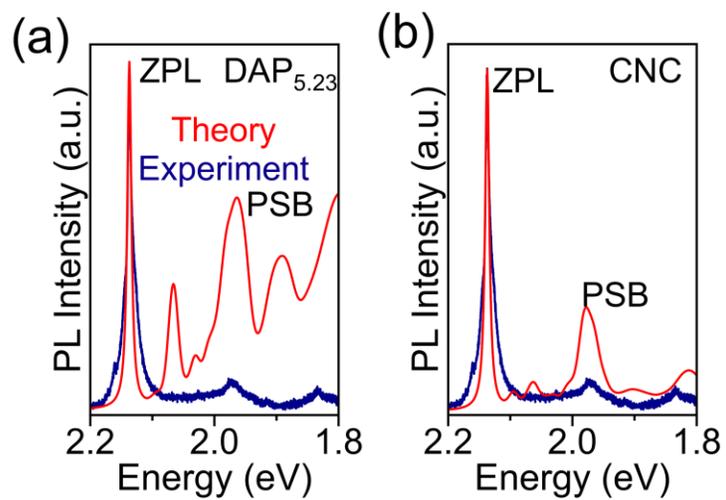

*Figure S14: Calculated Vs experimental PL spectra*. Calculated PL spectra (red) for the (a) DAP5.23 and (b) CNC. The calculated PL spectra are compared with the experiment (dark blue).

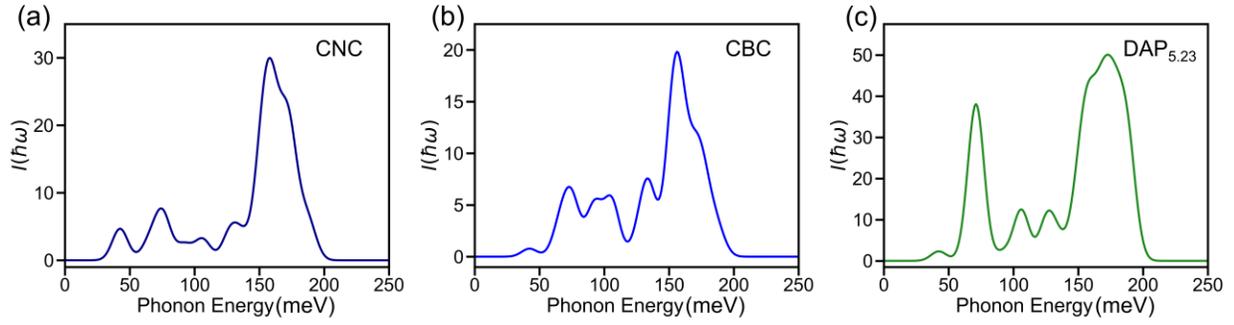

*Figure S15*: **Calculated phonon spectral function.** *Phonon spectral function calculated for (a) CNC, (b) CBC, and (c) DAP$_{5.23}$. The strongest peak is the LO phonon energy referred to in **Table 2** of the main text.*